# The Slow Space Editor : Broadening Access to Restorative XR

NATE LAFFAN, University of California, Santa Cruz, USA

ASHLEY HOM, University of California, Santa Cruz, USA

ANDREA NADINE CASTILLO, University of California, Santa Cruz, USA

ELIZABETH GITELMAN, University of California, Santa Cruz, USA

REBECCA ZHAO, University of California, Santa Cruz, USA

NIKITA SHENOY, University of California, Santa Cruz, USA

KAIA RAE SCHWEIG, University of California, Santa Cruz, USA

KATHERINE ISBISTER, University of California, Santa Cruz, USA

The Slow Space Editor is a 2D tool for creating 3D spaces. It was built as part of a research-through-design project that investigates how Virtual and Mixed Reality (XR) environments might be used for reflection and attention restoration. In this phase, we seek to radically simplify the creation of virtual environments, thereby broadening the potential group of users who could benefit from them. The research described in this paper has three aspects. First, we define the concept of "slow space," situating it alongside existing research in HCI and environmental psychology. Second, we report on a series of interviews with professional designers about how slow spaces are created in the physical world. Third, we share the design of the tool itself, focussing on the benefits of providing a simple method for users to control their environments. We conclude with our findings from a 19-person qualitative study of the tool.

## 1 INTRODUCTION

Over the past few years, the consumer XR market has seen a remarkable uptick in momentum. The release of Meta's Quest 2 and Quest 3, followed by Apple's Vision Pro represent tens of billions of research dollars and a bet that our primary mode of computing will soon be through XR systems. The growth of XR sales has followed a similar trend. In 2020, the consumer XR market was valued at USD 25.2 billion. By 2026, that number is expected to reach USD 361.9 billion[21].

Despite industry momentum, though, the broad, day-to-day adoption of XR is still on the horizon. We feel this creates an opportunity to learn from the lessons of the smartphone, which has been shown to be detrimental to our ability to regulate emotion [25], remember [22] and remain in control of our own attention [29].

Indeed, regarding this last point, there is compelling evidence that suggests we are experiencing a global crisis of attention. In his recent book on the subject, *Stolen Focus*, author Johann Hari frames how this problem can cascade: "Democracy requires the ability of a population to pay attention long enough to identify real problems, distinguish them from fantasies, come up with solutions, and hold their leaders accountable if they fail to deliver them. If we lose that, we lose our ability to have a fully functioning society. [...] A world full of attention-deprived citizens alternating between Twitter and Snapchat will be a world of cascading crises where we can't get a handle on any of them" [9].

Framed this way, it is easy to imagine a world where XR, if its software follows the same attention-grabbing paradigms as the smartphone, could result in even more severe outcomes. However, we believe that the qualities that give XR its particular power—total overwhelm of a single sense, a forced reorientation towards a fictive or semi-fictive environment—are qualities that could also be used to *restore* a user's attention, giving them a means to replenish their capacity for focus in an ever-accelerating world [24]. In short, we believe XR could be a tremendous boon to our capacity for focus and the maintenance of our attentional state.

In this paper, we consider this possibility from three somewhat separate vantage points. First, what does it mean for a space to be restorative? That is, what qualities of a space are demonstrated to be beneficial? Second, what design choices are available when building a space? That is, how do the exposed parameters of a tool aid the user in creating an intentional space? Finally, how does the user control the space? Are the interaction paradigms encouraging new users who might need these spaces? Or are they limiting space creation to existing XR enthusiasts? Considered together, these questions frame an exciting research space.

To engage with these ideas directly, we opted to use a Research through Design (RtD) process [30] and explored this conceptual territory using the design of a prototype. RtD is widely defined in modern HCI, so to be more specific, we are working with Zimmerman's summary of RtD's goals (expressed while summarizing the work of Christopher Frayling), which defines RtD as "[the creation of] artifacts intended to transform the world from the current state to a preferred state." [30] In this case, we are creating an artifact (the Slow Space Editor) which explores how personalized, restorative VR space creation might be made more accessible to those who may not be familiar with VR in general. The goal of this project was to develop actionable knowledge through the process of design and engaging study participants early to provide feedback and accelerate iteration cycles.

In the following sections, we will explain the thinking that went into this prototype. In the first section ( Defining Slow Space) we introduce the research behind reflection, green space and how it might be applied to VR - a practice we call building "slow" spaces. In the second section (Learning from real-world designers) we walk through lessons learned from a series of interviews with professional designers of physical space, discussing how they approach the design of such spaces in their own work. In the final section (The Slow Space Editor), we introduce the prototype itself, walk through its design process and discuss our findings from a 19-person user study.

## 2 DEFINING SLOW SPACE

### 2.1 Environmental Psychology and space-based behavior

The field of environmental psychology is premised on the idea that our spaces have a significant impact on our interior state, and, by extension, our behavior. This idea was first proposed in the 1950s through the work of Roger Barker and Herbert Wright, psychologists who developed the concept of "behavior settings"[17]. Their research showed that people's actions were more strongly predicted by their surrounding environment than by individual traits or immediate social influences[20]. These behavior settings are self-regulating, socially distributed patterns of activity that persist in specific locations and times, shaping and being shaped by the collective behavior of participants. In essence, they found that spaces trigger and help form different identities.

### 2.2 The benefits of green space

In the years since this connection was established, much research has gone into exploring the restorative power of space. By now it is now almost a truism to point out that that even limited exposure to green spaces (ie. nature) can result in a wide range of positive outcomes. In fact, not only is exposure to green spaces beneficial for mood [4] , but it can help reduce pain [23], restore attention [11] and even result in improved mortality [10].

Unfortunately, there are large populations who do not have ready access to such spaces. The readiest example of this comes from populations dealing with accessibility or health issues [28]. A study conducted by Browning et al. found that Americans spend over 500,000 days in hospitals every year, and in the United States and Europe alone over nine million adults live in assisted care facilities[5]. While some of these individuals may well be able to leave the building and enjoy green space, it is safe to assume that a large percentage of them are either unable to move through restorative space with the mobility they would like or are unable to access them at all.

Thankfully, while accessing these spaces in an immediate, physical way is best, there is now compelling evidence that we even benefit from a digital representation of the same. A recent study by Ünal et al. compared virtual simulations of urban and green space against their physical counterparts and found that the counterparts elicited similar effects, showing increased restorative characteristics in the simulated green space [24].

The best known theory of how these benefits function psychologically comes from Kaplan and Kaplan's 1989 work, *The Experience of Nature*. In it, they propose Attention Restoration Theory, which posits that exposure to natural environments can restore cognitive resources by possessing four primary qualities: being away, extent, fascination, and compatibility [11]. *Being away* refers to the sense of psychological or physical distance from one's routine activities and stressors, allowing for a mental break. *Extent* describes an environment that feels like "a whole other world" by emphasizing two interrelated characteristics: connectedness and scope. That is to say the elements of the environment should cohere, and feel that they "constitute a portion of some larger whole" (p. 184). *Fascination* refers to elements in the environment that effortlessly capture attention. (This is perhaps the most nuanced requirement, as there are pitfalls to be found on both sides of the equation: an environment that is too fascinating runs the risk of simply being another distraction, whereas static environments are dull, and run the risk of under-stimulation.) Finally, *compatibility* means that the environment supports the individual's intentions and goals, requiring little effort to adapt or fit in.

## 2.3 Territories

In addition to the value of green space, environmental psychologists have also found that a person's sense of control or ownership over a space can aid in its restorative value. The term used to describe these spaces is *territories*. Research on human territories, such as home environments, has shown that physical settings can guide behavioral, cognitive, and emotional processes[17]. Territories become synomorphic (ie. linked between environment and behavior) with patterns of activity as "residents" spend time exploring, designing, and inhabiting them. This familiarity allows for more efficient navigation and, consequently, a reduction in stress. Crucially, territories satisfy a fundamental psychological need by allowing residents to express their identity through marking behaviors and environmental alterations, referred to in the literature as *self directed identity claims* [17]. This process shapes both the environment and the resident's self-concept in a symbiotic cycle.

## 2.4 Restoration and HCI

As mentioned above, there is evidence to suggest that we may be in the throes of an attentional crisis, potentially exacerbated by digital tools. Yet the question of how digital tools might be used to *benefit* our attentional state has been an ongoing conversation within the HCI community for decades. In 2001, in the paper *Slow Technology - Designing for Reflection*, authors Hallnäs and Redström rightly predict that "when computers become increasingly ubiquitous, some of them will turn from being tools explicitly used in specific situations to being more or less continuously present as a part of a designed environment," and that we'll "need actively to promote moments of reflection and mental rest in a more and more rapidly changing environment" [8]. The authors assert that it is time for "a design agenda for technology aimed at reflection and moments of mental rest rather than efficiency in performance"[8]. They refer to this agenda as "slow technology", and it is here, within this frame, that we locate our work.

## 2.5 Definitions

What exactly is meant when Hallnäs and Redström write "reflection and mental rest"? In seeking to better understand this question in the virtual sphere, we have broadened our search to include fields from the physical sphere, and in the process have become obliged to use terminology as it is understood by each. For example, during our literature review, we found that despite its frequent use in the world of HCI, "reflection" is not a term commonly found in architecture literature. Similarly, much appears to be written about "contemplation" as it relates to landscape architecture, but little can be found about supporting it in HCI scholarship. However, there is an important link. In both cases and across fields, the stated goal of these activities is often framed in terms of restoration. With this in mind we will briefly touch on what definitions are being used.

*2.5.1 Reflection.* Definitions of reflection, frequently invoked in HCI papers as a positive end state or goal, cover a great deal of conceptual territory. As Baumer et al. shows in their review of the topic, these definitions often align with broader contextual goals, and tend to orient around education, design or self knowledge; having roots in the work of professional or educational theorists. Rather than recapitulate the finer points of this investigation, we will rely on Baumer's definition directly—a mixture of Schön, Dewey and Moon's—which defines reflection as "reviewing a series of previous experiences, events, stories, etc., and putting them together in such a way as to come to a better understanding or to gain some sort of insight" [2].

*2.5.2 Contemplation.* The literature on reflection tends to emphasize studying, and the concerns that crop up around the presentation of information. Consequently, we have found there to be scant focus on the broader context in which the reflecting is taking place. However, by pairing this word with mental rest, Hallnäs and Redström suggest a calm, peaceful environment. In landscape architecture, when designers are creating spaces for this kind of inward-focussed attention, we have found the word most often used is *contemplation*. Using this word as a guide has proven fruitful, and in this paper, we follow Olszewska-Guizzo's lead in using the Collins English Dictionary definition of contemplation: "an act of intentional, attentive watching, the perceiving of something, or thoughtful observation" that "induces and restores positive emotions, reduce stress and mental fatigue" [19].

In this paper, we use these two concepts - reflection and contemplation - to describe two different forms of attention: focussed and diffuse. In both cases, restoration is not an inevitable outcome (rumination being a negative practice which can come from either) but simply the one which we are seeking to support.

*2.5.3 Restoration.* The concept of restoration, even within the bounds of environmental psychology, is also widely defined. In the world of architectural planning, Zhang et al. frame the restorative environment in terms of productivity, defining it as "the capability to reduce mental fatigue, improve productivity, and relieve stress" [14]. Marcus and Sachs, whose work revolves around therapeutic gardens, define it in terms of the balance of positive and negative emotion, suggesting it is a "reduction in negative feelings such as fear and anger/aggression and improvement in positive feelings" [16]. Both definitions are missing a critical component, though, which is the acknowledgment of a previous state to which the individual is trying to return. Ulrich, perhaps the best known researcher in this area, emphasizes this, pointing out that restoration is "a broader concept that is not limited to stress recovery situations, or to recovery from states characterized by excessive psychological and physiological arousal, but could also apply to recuperation, for instance, from under-stimulation or excessively low arousal" [23]. As we will see later on, this notion of under-stimulation or low arousal is often obscured by concerns regarding overstimulation, and is an important area of concern when designing for XR. In this paper, we combine these three definitions and define restoration as "recovery from over or under stimulation that reduces mental fatigue and relieves stress."

## 2.6 Slow Space

Taken together, these three goals - reflection, contemplation and restoration - combine to form the basis of what we refer to here as *slow space*. Slow spaces are immersive, but not tied to any one technology. Although this paper concerns itself with the construction of slow spaces using XR, there is no reason that such spaces could not exist outside of that context and support the same fundamental goal.

Slow spaces are able to support focused or diffuse forms of attention. That is, both reflection and contemplation. They accomplish this by surrounding the user with an environment that does not demand attention, but rewards attention if it is given. As we will show later on in the paper, the most direct method of achieving this is by mimicking nature itself, and, while this is not a requirement, many of the ideas built into the Slow Space Editor are based off of this technique.

Given the potential for fantastic or impossible environments in XR, slow spaces are not bound by the tenets of evidence-based design, but the suggestions we have collected here largely draw from that world, as we believe there is a significant overlap between the way spaces in the physical and virtual worlds affect our capacity to direct attention. As such, they are concerned not just with the visual environment, but the acoustic one as well.

In many ways, slow spaces exist to counteract the effect of fast spaces. To adapt Hallnäs and Redström's description of fast technology, fast spaces emphasize "efficiency in functionality with respect to a well-defined task" and their general aim is "to take away time." [8] In contrast, slow spaces are design to extend time - to slow it down to a pace that is more humane, and which may run counter to logic of efficiency or productivity.

*2.6.1 Divergence from "Slow Technology".* While the *Slow Technology* paper was the inspiration for our development of slow spaces and has clearly informed its framing, it is important to note that there are some major differences between the two ideas. The most significant of these is our focus on restoration, and our understanding of the concept of reflection. As Hallnäs and Redström see it, slow technology "is not supposed to reduce cognitive load" [8]. While they intend the user to reflect, their aim is for the technology itself to be exposed and become the object of the user's reflection. Slow Technology, they write, should also "focus on aesthetics of material and use simple basic tools of modern technology." Here too our thinking differs. As we have noted above, slow spaces need not be mixed-reality systems, or even used head mounted displays. However, the aesthetics of the technology are less important when the goal is to minimize technology's conceptual presence rather than make it a focal point.

These are fundamental differences, and it is important to note that it is likely that Slow Space, though drawing on the same fundamental set of arguments as Slow Technology, likely does not qualify *as* Slow Technology, at least as it was originally envisioned.

## 2.7 Related Work

Given how well suited VR is positioned to provide restorative spaces to those who need them, it comes as no surprise that other research has investigated similar territory. There has been much valuable work investigating the value of VR as an aid to meditation [12, 18, 26], self-regulation [27] and enhanced attention [13]. And indeed, the researcher Simone Grassini, in a recent paper entitled *The use of VR natural environments for the reduction of stress: an overview on current research and future prospective* has done important and exhaustive work chronicling the many examples of research that exists in this space [7]. (Two recent examples stand out in particular. *Zenctuary VR: Simulating Nature in an Interactive*

*Virtual Reality Application* [1] and *Designing Virtual Natural Environments for Older Adults* [15].) We hope to distinguish ourselves amidst this ongoing work by framing it in the context of Slow Technology and, in subsequent sections, placing particular emphasis on the *creation* of such spaces, such that their benefits may reach a broader audience.

Apps in this category are often geared towards meditation, seeking to support focus or calm. During the writing of this paper, we investigated several recent apps, Maloka, Mindway, Project Flowerbed (Meta), Realms of Flow and Flowborne VR being particularly representative examples among them. In each case, we found that all fall into one of the following design paradigms: (1) 360 degree video, which rewards attention but isolates the participant from their surroundings [3], (2) low-poly modeling, which provides a high degree of relational presence but does not reward attention (discussed below) or (3) absence of natural elements, which run directly counter to the findings already discussed.

Exceptions do exist. In particular, the most recent versions of Nature Treks VR, Brink XR and Kayak VR all appear to qualify (to great or lesser degrees) as slow space. Indeed, the one area of development where these findings seem to be taken into account is gaming. This is not a surprise, as the gaming industry tends to operate at the cutting edge of immersive environmental design. However, although there are certainly games which seek to nurture mental rest, we have again not found any literature which draws connections between the design of those games and the principles we list below. A goal of this paper is to draw those connections explicitly, and suggest the value of constructing these spaces outside of a single category of app.

## 3 LEARNING FROM REAL-WORLD DESIGNERS

Early versions of the prototype led us to an important (if obvious) realization: designers who have created these spaces in the physical world may have useful insights that designers of virtual space could draw on. It was with this in mind that we arranged a series of interviews with design professionals about their process of designing spaces for reflection or contemplation. We interviewed eleven designers who came from a broad range of fields, including architecture (5), landscape architecture (3), horticulture (1), lighting design (1) and interior design (1). The conversations were not highly structured. Instead, they revolved around three main themes: background information about how the individual rose to the position they have now, what the design process in their particular field looks like, and finally, how they might think about the construction of a contemplative place, both in the real world and in XR.

These interviews were conducted entirely over Zoom, and lasted anywhere from 30 - 70 minutes, depending on the interviewee's availability. As each expert acted within their professional capacity and did not request anonymity, we will refer to them by their full name. After careful (informal) analysis, we came away with three core areas to focus on as we began to construct the editor.

*3.0.1 Iterative prototyping is key.* Many interviewees described their design process as highly iterative. Their approach involved multiple site visits, hands-on modeling, and direct client interaction in the space. Matthew Girard, a landscape architect, emphasized the importance of physically mapping out designs on-site: "We'll flag it out [...] and lay it out on site to see what we've drawn and how it works." Mr. Girard stressed that this back and forth between the design and the space being designed is crucial: "It's truly the only real way to know what you're making." Andrew Barnett, an architect, echoed this when he highlighted the value of physical models in the iterative process: "I like models because they can be less resolved in a way, and you can easily change them." He noted that physical models allow designers to "investigate just the certain element that you're looking for [...] and ignore everything else." Another architect, Eirini

Karamolegkou, agreed, pointing out that in her process "we work a lot with models in a way to really test our initial ideas in terms of height, form and views."

*3.0.2   The value of realistic dynamism in nature.*  Nearly all interviewees highlighted incorporating nature and natural elements as crucial for reflective or contemplative space. Ms. Karamolegkou emphasized the importance of "a connection to nature" in designs, saying elements like "natural light and landscape views" are "fundamental to our approach to architecture". Julie Moir Messervy, a landscape designer and author who specializes in the design of contemplative space, said contemplative gardens should have "movement" from elements like rustling leaves, which helps "detach the mind from the physical." Donna Brown, an architect, points out that the dynamics of nature "make you feel connected to the bigger world. And your challenge in creating a virtual environment is that it's not an isolating environment"

Hallie Schmidt, a professional horticulturist and garden designer, emphasized how critical detail can be. She is worth quoting at length on this point: "In contemplative space you're drawing connections and creating ideas and thinking about things. Those are inspired by a lot of what you see in nature. [In VR] at a certain point you've created the elements and the shapes [...] and you're sitting in that space in VR and now you've seen it all. And you're like, okay, there's a tree, there's some grass. What else is there? What's interesting about this? The amount of complexity that exists in reality, where even on a blade of grass, [...] you can see where the insects have nibbled on it, you can see where it's died back a little bit at the tip. [...] You can really get up close and there's so much complexity. In VR, I'd imagine it's just a green blade."

*3.0.3   Design for exploration.*  Given the emphasis on complexity, the concept of "mystery" in landscape and architectural design is closely aligned with the notion of extent. Mystery evokes a sense of curiosity and invites exploration, much like how complexity in an environment engages attention. An environment with "mystery" offers partial or obscured views, encouraging the observer to move forward and discover what lies beyond. Mr. Girard has worked on gardens which adopt these techniques. He points out that "when you look into a landscape, it is dead if it is revealed to you at once. But if there's something that's in your view - a boulder, a hill, a dense evergreen - and you're thinking, oh, there is something around that that I can't see, then you feel a connection. It's where you feel you're drawn."

## 3.1   Designing towards "worthiness"

After the interviews concluded, these three themes - rapid prototyping, realistic/dynamic elements and exploration - became the cornerstone of our design process for the Slow Space Editor. However, the current state of graphics processing renders some of these considerations more feasible than others. Realistic dynamism is a serious challenge for virtual environment designers, regardless of their industry or focus. Increased complexity translates directly to processor load, and to keep frame rates high enough to prevent nausea, today's solutions often use cartoonish abstraction to lighten the computational requirements of a scene. Consequently, they tend to elicit exactly the experience Ms. Schmidt describes: being merely *reminded* of an object that is worthy of our attention.

This is not to say realistic dynamism should be abandoned. As apps such as Nature Treks VR, Brink XR and Kayak VR make clear, the boundary of what is possible is constantly moving forward. Though we had neither the expertise nor the budget to pursue such fidelity in our own tool, we think it worth noting that this "worthiness" of attention should be considered a cornerstone of slow space design.

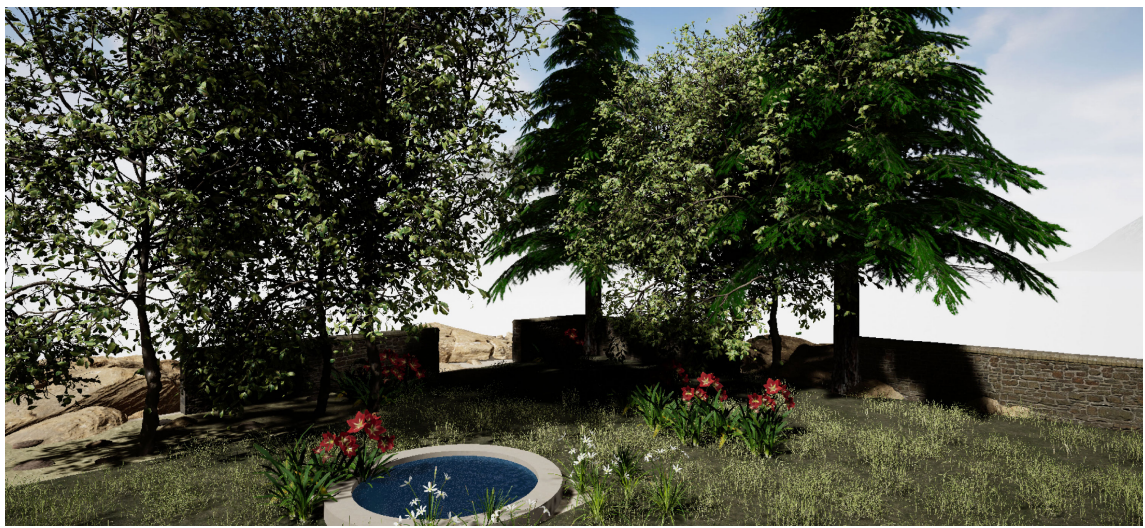

Fig. 1. An early demonstration garden, used for recruiting study participants.

## 4 THE SLOW SPACE EDITOR

These interviews gave us a clear direction as we began working on the prototype. We started the design process by setting out the above lessons as design requirements. The tool would need to support rapid prototyping, allowing users to quickly adjust course as they designed their space. The environmental tools that users are given to work with should emphasizes green space, possibly geared towards existing gardening conventions. Moreover, these green tools should in some way be dynamic and believable, hewing as closely as possible to the kind of detail that makes attention restoration in physical nature possible. Finally, in should encourage exploration, either of the space itself or the space of possibilities.

### 4.1 Tools

After an initial round of experiments with WebXR, we decided to switch to Unreal Engine (version 5.2) due to its ability to produce a nearly photo-realistic experience when using an untethered Meta Quest 3. Unreal also provided us the opportunity to try out different methods of rapid prototyping, thanks to their large library of high-resolution models in the Megascans library.

### 4.2 Environment sketches

Before constructing the tool in earnest, as part of our RtD process, we decided to explore our concept of slow space through a series of "environment sketches" (see Figure 2). These were a series of quickly developed spaces inspired by members of the research team imagining their ideal restorative setting. This was a particularly fruitful process, as it became clear that not only were the spaces we created effective for the person who had designed them, they also could be shared as an artifact unto themselves, by inviting others to try on the headset when the app was running. This created several formative conversations about ideal places to think. These sketches confirmed not only that Unreal was a good choice of software, but that the rapid prototyping of spaces, once simplified, could potentially be a social activity as well.

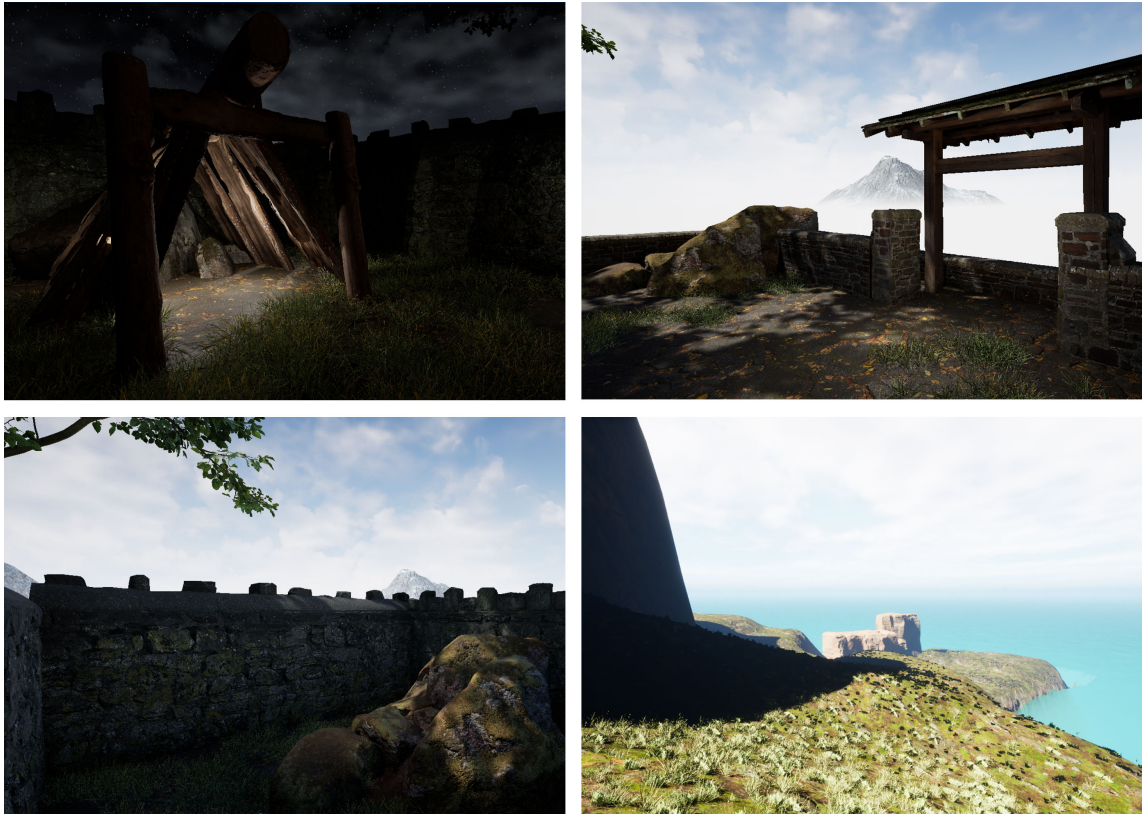

Fig. 2. A series of "environment sketches" were created using Unreal Engine to explore individual ideas of restorative space.

### 4.3 The Editor

The Slow Space Editor (see Figure 3) has three main entry points : an editor (2D), a preview window (3D) and a VR experience (3D). The editor is a simple web interface built using the Phaser game engine. As changes are made in the 2D editor, the 3D spaces are updated in real time using the WebSockets protocol. The editor's main area consists of a two dimensional grid that is mapped directly to the three dimensional space in the preview window. Clicking the lines of the grid creates segments of wall, allowing the user to close off space. Clicking on a square of the grid cycles that segment through grass, rock and water textures in the 3D spaces. Along the side of the grid are a series of icons, representing the items that the user is able to instantiate in the 3D space. When the user drags an item onto the grid, it instantly appears in the preview and VR spaces. Placed items can be deleted (by dragging them into the trash) or moved around the grid. In the upper right hand corner is a sun icon, which the user can click to cycle through different times of day: morning, dusk and night.

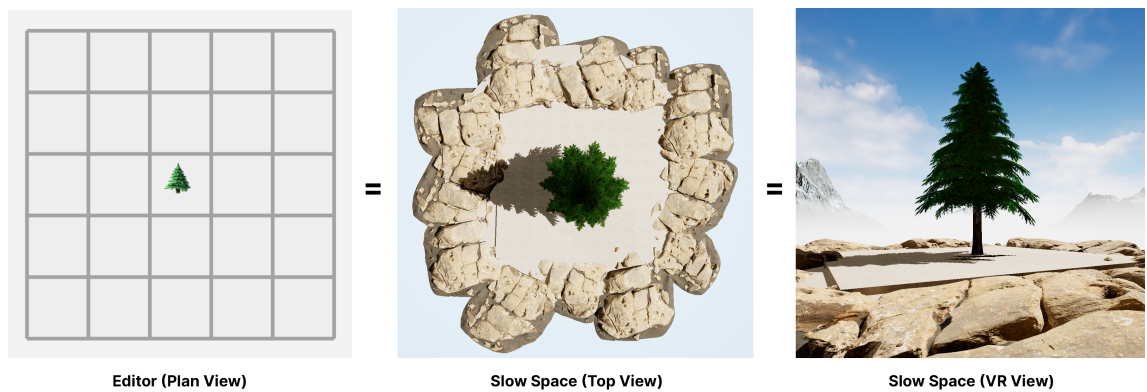

Fig. 3. These images show how the 2D plan view of the editor relates to the Slow Space itself. On the left is the 2D interface with a single item added. At center is how this item would look from above in the 3D space, and on the right shows what the user can expect to see when putting on the headset.

## 4.4 Radically simplified editing

The decision to use a 2D interface to create the spaces (rather than provide users with a method of controlling items directly within the environment) was born of three, somewhat interrelated considerations.

First, the editor must permit a coherent understanding of the *entire* space to allow for rapid prototyping. In professional 3D editing software, this comes in the form of top and side views, providing the user a clear picture of how the pieces fit together. In VR, however, there is rarely this abstracted view. Recalling a theme from our interviews, the schematic view we provide is not unlike the use of models - it gives the user the ability to relate to the design in its totality, not just the aspects of it which they are immediately surrounded by.

Second, it must be simple enough to engage even those who are unfamiliar with VR. Although VR tools such as Tilt Brush, Gravity Sketch and ShapesXR have radically simplified the experience of creating 3D models in VR, users must still master the navigation of a complex set of menus or gestures in virtual space. For younger users or existing enthusiasts, this may pose little challenge. However, within the groups we hope to reach (ie. those unable to access green space due to mobility issues) would likely be a high number of older users or those who are unfamiliar with VR. A 2D, drag-and-drop editor requires no new skills, and the user is therefore able to create within VR's extraordinary immersive space with significantly less cognitive overhead. We decided to use emojis for the icons as, given their use in smartphone interfaces, users would likely be familiar with them and they are designed to be immediately recognizable at a small scale.

Finally, after spending many hours editing spaces within VR during earlier projects, we came to the conclusion that although VR headsets are lighter than ever, they are not yet comfortable enough to assume that every activity related to their use should be executed inside them.

To be sure, this approach deviates from the norm, creating a simulation gap between the two corresponding environments. However, we feel that this cognitive cost is far outweighed by the benefit of familiarity, where users who may not be familiar with the affordances of VR are able to control a VR space through a common interaction modality like drag-and-drop. In addition to wildly simplifying development, this also allows for spaces to be co-created in real time, which, though not the focus of this study, we think could be an exciting direction to take this format.

## 4.5 Study Methodology

As discussed above, throughout this project we followed a Research through Design (RtD) process [30], aimed at creating a prototype through which we could explore a possible future. In this case, that future was one where Slow Spaces could be built in VR by users with no VR experience.

After an early pre-study testing round with members of the research team, we recruited 19 participants to use the tool in a lab setting. Given the exploratory nature of the prototype, we used a convenience sample, primarily composed of available undergraduates and friends. Participants ranged in age from 19-40, with the median age of 21.8. Almost all had a technical background (n=18) and many had used VR before (n=14).

During each session, the participant was given a brief introduction to the tool and then asked to "pick something you've been meaning to think about and build a garden to think your thought." During the construction of their garden, they were encouraged to use the think-out-loud method to help reveal the rationale behind their decisions. Once the garden was completed, they were asked to spend two minutes inside it using the VR headset. Once the two minutes were up, a researcher asked them a series of questions about their experience.

After the conversations had been transcribed, each researcher made a first-pass coding of each conversation, including memos of particularly salient or valuable phrases. These codes and memos were then discussed during a weekly researcher meeting, the notes from which also became part of the dataset. Once all the data had been coded, it was used to create a codebook with the help of a large language model (Anthropic's Claude Sonnet 3.5), which facilitated the surfacing of primary themes and quotes. From these themes and quotes we assemble the following observations.

## 4.6 Observations

*4.6.1 Ease of use.* As a core goal of this project was to provide an environment creation tool to individuals with no XR experience, we placed special value on the transcripts of the five participants who fit this description. Their responses, captured just minutes after trying VR for the first time, were promising. Participant 1 (P1) described the experience as "super easy to use, at least for me. And fun." For P3, it recalled other world building experiences from their childhood : "It was nice. It kind of reminded me of when I was younger. I built Lego worlds. It just kind of reminded me of that. Like, I can create my own, own little world with everything that I want, in the place that I want it to be." P11 described it as "a pretty good, straightforward, intuitive system that allows you to place something and get a good feel for where everything goes."

However, others found the tool to be too stripped down. Several participants with XR experience expressed a desire for basic manipulation features of the sort one might find in a typical 3D editor, such as rotation or scaling. This extended to the palette as well. One participant felt that it was well suited to the task, saying "I think that it's nice to have not a ton of options" (P19), while another felt differently, pointing out that "I had to build the environment around the objects instead of building the environment the way I would" (P8).

*4.6.2 Safety first.* Despite the study taking place in a fully enclosed room, participants frequently created spaces that emphasized safety, privacy, and comfort for their physical self. Participants often built their "safe spaces" first, even before trying on the headset. As one participant put it, "I'm very conscious that I don't want to block the view. But I also want it to feel safe. So some nice boulders on the edge will make me feel safe." (P19) Another pointed out "I kind of wanted it to be surrounded by trees. Just so it felt closed off. You know what [walls] signify in a park. They make you feel safe and enclosed in a space" (P7).

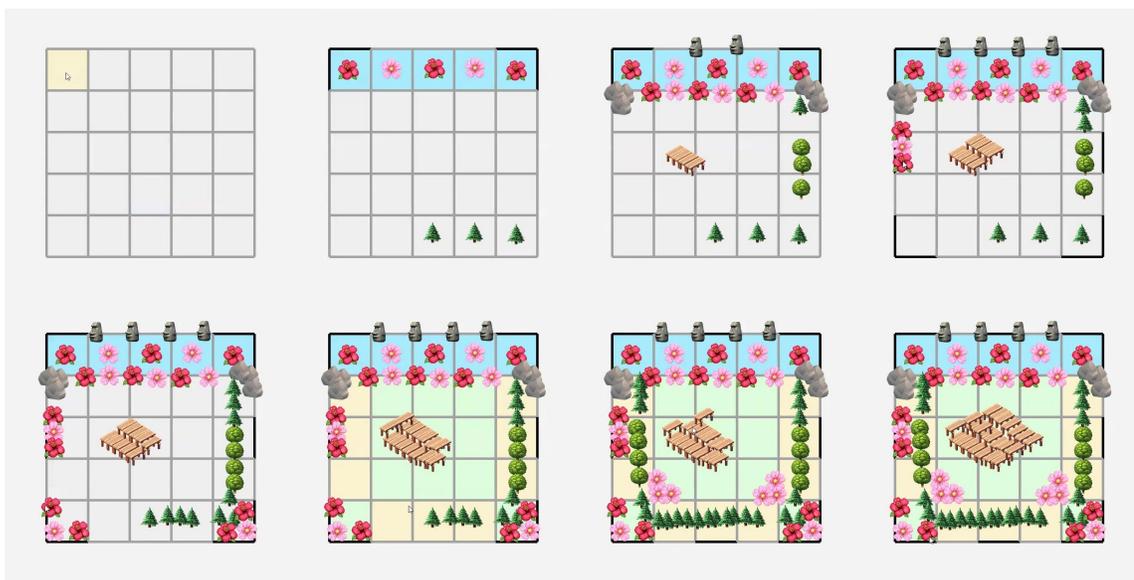

Fig. 4. A participant recreated a familiar spot on campus.

*4.6.3 Virtual spaces for physical bodies.* These safety considerations were often mixed with a broader sense of designing for bodily comfort, even though their physical body would never occupy the space. One participant explained a series of floor tile choices by saying "if you wanna take your shoes off, you could feel the grass and then where all the trees are, there's like sand and gravel" (P16). Another participant, when describing the placement of a bench, said "I want it to be wide enough to be comfortable to sit on. I want the whole area to be grassy so it's comfy to walk on" (P7, see Figure 4).

Although for the most part these choices were explained by actions that could *not* be taken, two participants build spaces that they would not have been able to experience in real life. As one of them explained "I think what's fun is that I'm actually allergic to flowers. So I don't actually like go around them, but because it's VR, I can just stack a lot of flowers knowing I'm not going to get a pollen allergy" (P6).

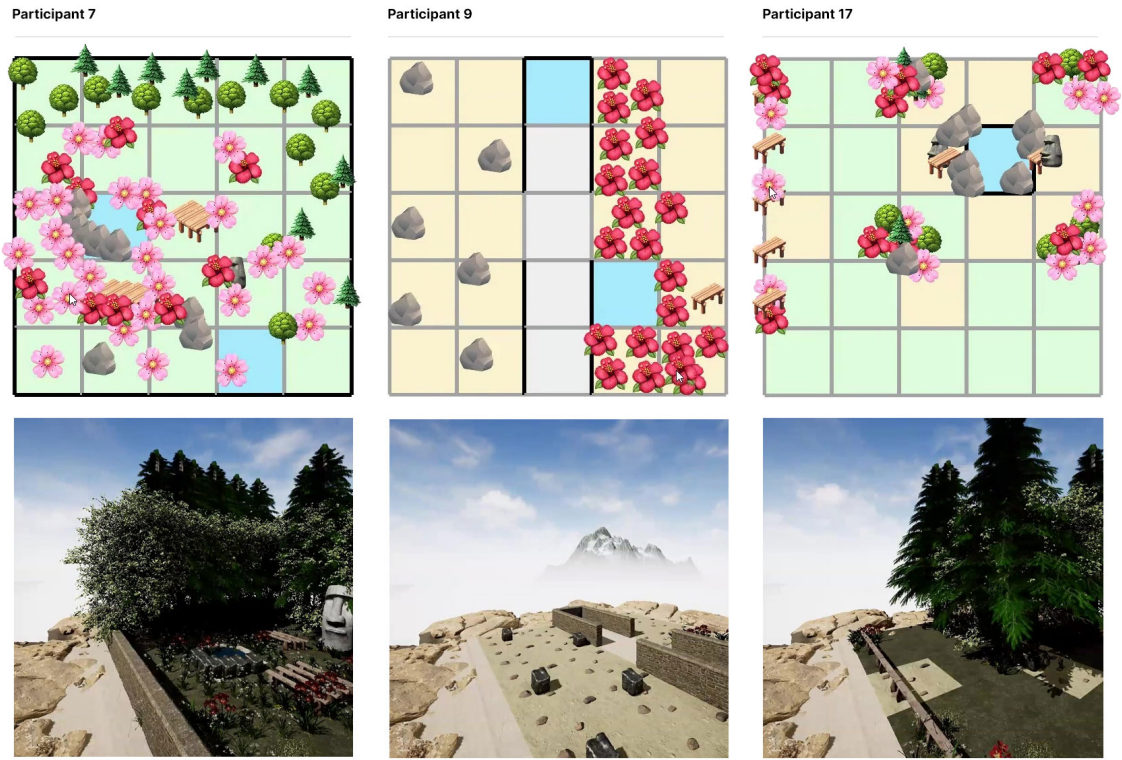

Fig. 5. Several participants told stories about the spaces they were building.

*4.6.4 Prototyping = storytelling.* Most striking to the research team was the degree to which participants could, after less than a minute of instruction, use the tool to sketch stories from memory, ideate scenes or even problem solve. (See Figure 6.)

One participant built the view from their grandmother's kitchen, explaining - "[this is] the main artery of the town. I think I mentioned that the town is built around a single road as it leads out from a larger road, a highway. And on one side there was at least the way I remember it as a child, there was the livable part" (P9). Another found this ability to be impactful on its restorative qualities, noting - "I kind of built it with my childhood in mind. [...] I think being able to like recreate like somewhat close to memory definitely made it more effective" (P7).

Other participants worked out elaborate stories of use, despite their only directive being "a garden to think your thought." For example, one participant began to describe the beginnings of what sounded like a public park : "kids got to be able to hang out over here and be chill and cool. So I made a little sandpit with some water so they can build, like, a big sandcastle. And then they have this guy watching over them. And then this bench over here is for the parents to look over and watch the kids" (P16). Another participant built a garden that was itself a representation of reflection: "I decided on there as a kind of symbolic thing. Both of [the statues] are in individual contemplation. The one right here is looking into the well because he's contemplating himself, and the one right here is looking outward because he's contemplating the well" (P17).

*4.6.5 Dynamism matters.* As might have been predicted from our conversations with professional designers, nearly every participant suggested that increased dynamism or surprise would have made the environment a more pleasant place to spend time. This cut across many aspects of the design, and was framed many different ways. For some, this could be as simple as audio changing with each addition to the scene. One participant pointed out that "something that I usually feel like I need for areas to feel calm is the sound of nature. […] I love hearing the water rushing or like birds and the wind" (P14). A common refrain was the need for some kind of animal life to inhabit the scene. As one participant pointed out, "[in nature] I'd see more life, like animals. So if there were birds or squirrels in the trees, then that would feel more realistic" (P1). Another, when asked what changes could be made to the tool, suggested "I think maybe like any moving element. When I was staring at the pond. I was thinking like. Oh, like, what if there were koi fish or something in it?" (P14)

This need for dynamism came out in other, more subtle ways as well. Seeking to *create* surprise, one participant used the walls to construct a series of spaces that he could walk through, explaining "the idea is to add enough areas where once you go through all of them, you kind of forget what one of them is and you're like, oh, you know, let me revisit this one" (P16). Other participants suggested methods for making even the static options feel more dynamic, imagining that in a future version objects would be "more random instead of just static. You have like, ten rocks and it just picks a random rock instead of the same rock over and over" (P6).

| Participant | Quote |
| --- | --- |
| 001 | "Yeah I would, but there's no there's nothing to do." |
| 003 | "Yeah. Yeah, yeah." |
| 004 | "Yeah, I think that would be kind of neat." |
| 005 | "Yeah" |
| 006 | "Yeah, I would." |
| 007 | "Yeah, I think so. I would." |
| 008 | "Um. If it was like not a long hike there. It's like, this looks like it's on a mountain or something. But like, if I could just, like, instantly pop them into there and just show them." |
| 009 | "No." |
| 010 | "Maybe. Yeah, I'm pretty social." |
| 011 | "Yeah, I think that would be kind of neat. It feels like a little empty and lonely the way it is." |
| 012 | "Probably not. I feel like this is more of, like, a safe. Like a safe place just for me." |
| 013 | "I would want to bring my girlfriend in." |
| 014 | "I think so. I think with, like, the intent of like having quiet time. Yeah, like studying or something. This would be a really nice spot to just kind of, like, step away from everything." |
| 015 | "Yeah, I would." |
| 016 | "Um, I mean, sure, I would invite all my friends because it's a lovely area and I think it's great." |
| 017 | "Yes? Obviously we would be eating lunch in the space because virtual food doesn't do anything." |
| 018 | "I don't know, that's a good question. I feel like I'll leave the space more just for me to think about things and kind of away from other people." |
| 019 | "Absolutely." |

Fig. 6. Each participant's response to the question of whether they would like to invite a friend in to the space.

*4.6.6 Social potential.* Each participant was asked if they would invite a friend into the space they had created. Their responses were striking (see Figure 6). Some preferred the space to be private, but the overwhelming majority responded that they would invite a friend, and in one case they even had someone already in mind. Given the project's emphasis on personal reflection, this arrived as a surprise, but as one participant explained "it feels like a little empty and lonely the way it is" (P11). Another participant expressed a desire to see "community made maps." They explained that this would mean that "people were able to make or create whatever they wanted and post it somewhere and be like, oh, this is what I created. Look at this" (P16).

## 4.7 Discussion

In the next section we briefly discuss limitations of the study and then propose four potential areas which seem particularly ripe for further exploration.

*4.7.1 Limitations.*

*Sampling.* Given that this was an early-stage prototype, and our tests were focussed on usability issues and iteration, we relied on convenience sampling. This permitted a more rapid testing cycle, and was effective in demonstrating the efficacy of the concept. However, given the groups who we think would benefit most from the Slow Space concept (ie. those with little-to-no VR experience and without access to green space) future work in this area should recruit participants from more representitive demographics.

*Realistic Dynamism.* As mentioned above, we feel strongly that when designing for restoration, realistic dynamism is a critical factor. However, given our skill level within the Unreal development environment, the items that users placed were largely static, and therefore we could only gesture at the kind of experience that is possible today. As a prototype, we feel this was successful in validating our other areas of interest, but future work would need to to make realistic dynamism a focus before a more in-depth study could be conducted. (See potential next steps in the "Evidence of Time" and "Automatic Variety" sections below.)

*4.7.2 Social spaces.* There are many ways in which this tool, currently focussed on the individual, could be made to support more social interactions. Because the editor itself exists outside the virtual space, methods less commonly used in VR present themselves as interesting options. For example, editing a space could be an in-person collaborative activity. Two people could edit a single space (that is revealed in two headsets), or one person could edit while the other gives commands from the headset. Or, given that the editor is already configured as a website, editing could be an online collaboration, much the same way that modern web tools allow text documents to be edited by many users simultaneously.

*4.7.3 Personalization.* As the literature on territories points out, territories are not only familiar, they can become central to an individual's self concept [17]. It would therefore make sense to allow individuals to bring personal items into the space in much the same way that professionals sometimes bring photographs to their desk to remind them of family, friends, and identities that exist outside the professional sphere. In his ecological account of territories, Meagher even suggests that the more a territory is associated with one's self, it may be more likely to elicit soft fascination in a way that is potentially restorative [17].

*4.7.4 Evidence of time.* In a similar vein, several of the designers that we talked to mention the passage of time as being a natural detail they sought to highlight with their designs. Interestingly, the natural passage of time is somewhat overlooked in the examples we investigated for this project. But there may be real value in suggested continuity between one's actual physical season and what is waiting in a virtual space. On a smaller scale, a more obvious version of the same would be a method for exhibiting a user's use of the space - a method of virtualizing the wear and tear that occurs in an individual's physical territory. This passive record of use (referred to in the literature on territories as *behavioral residue* [17]) might make the space feel more closely tied to an individual's identity.

*4.7.5 Automatic variety.* Procedural content generation (PCG) could create a bridge between the simplicity of the editor and the complexity required in the environment itself. Although there is currently a 1-1 relationship between what is seen in the editor and what is seen in the environment, we can easily see the value of having the editor be abstracted somewhat from the components of the scene. For example, instead of a tree icon simply placing a single tree, one can easily imagine it setting the position of a node within a larger procedural content graph. In this scenario, rather than a static tree being placed, the node could instantiate a unique ecosystem that contains the details and dynamic elements that make the tree "worthy" of attention.

## 5 CONCLUSION

In this paper, we used a Research through Design process to create the Slow Space Editor, a prototype that helps to investigate how XR environments might be used for reflection and attention restoration. In the first section, we introduced the concept of *slow space*, as it relates to reflection, restoration and contemplation and found compelling evidence that such spaces should be composed of natural elements and be controllable by the user. In the second section, we introduced insights from the designers of physical space, which led us to focus on prototyping and exploration as key considerations in the design of the tool. Finally, we described the process of designing the editor, explained its basic functionality and reported on a brief user study, which showed there to be exciting potential in the radical simplification of space-creation tools. Most notably we found that users grasped the potential of this tool with almost no instruction, and used it in ways we could not have predicted.

In the introduction, we argued that the groups most in need of slow space may be those who have no access to it in the physical world. This raises an important ethical question that is worth touching on as we conclude. The potential dominance of XR, especially as it relates to attention and our ability to engage meaningfully with each other through the shroud of mediated space, is a contentious issue. Indeed, dystopian fiction often deploys VR as a symbol of humanity's disinclination to acknowledge hard problems or the ways we have become enslaved by our own innovation. These symbols feel especially salient now, as we enter what many are describing as the *anthropocene* - an era of human activity that is significantly and detrimentally affecting our climate and ecosystems. Is slow space not an attempt to anesthetize our response to disaster? Does it not just digitize what we are actively destroying elsewhere? By invoking "those who may not have access," are we not taunting those who were robbed of their contemplative space in the first place?

As we engage with this question of what role XR should play in the future it can feel as though the stakes could not be higher. But, at the risk of sounding naive, we both acknowledge these criticisms and maintain that slow spaces could help alleviate the problem, rather than exacerbate it. By aligning our introduction of slow space with theories of attention restoration, we hope to stand in direct opposition to the now well-known dark patterns that have been so ruinous to our ability to willfully direct attention [6]. By taking the best practices of space-building from the physical world, and by using what we know about how our environment can impact our own psychology, we believe these spaces could play an important part in restoring our ability to contemplate and reflect deeply. Ultimately, we feel that it is this kind of attention, where we are given space to ponder and come to our own conclusions, that is the route to our best selves.